\documentclass[12pt,a4paper]{article}

\begin{document}
\title{Some Discussion on Thermodynamical Behaviour of Modified Chaplygin Gas}

\author{Balendra Kr. Dev Choudhury \footnote{E-mail: bdevchoudhury@yahoo.com}\\
Julie Saikia \footnote{E-mail: jsaikia2000@yahoo.co.in}\\ 
Deptt. of Physics, Pub Kamrup College\\
Baihata Chariali - 781381, Assam (INDIA)}
\date{}
 
\maketitle

\begin {abstract}
\noindent On going good number of research works establish Modified Chaplygin Gas as one one of the most favoured candidates of Dark Energy. In our present work, new bound on the parameter space of associated model - parameter has been worked out with new explanation. Moreover, the explicit relation between the mysterious state parameter $w$ and scale factor $a(t)$ has been derived.
\end {abstract}

{{\bf{Keywords}:} Chaplygin Gas, Dark Energy, Generalised Second Law (GSL)}

\section {Introduction}

\noindent Modified Chaplygin Gas is one of the most favoured candidates of Dark Energy. Though on observational ground the existence of Dark energy and Dark matter is almost confirmed [1,- 6 ], their characterizing natures are still in dark. So far, only we know the Dark matter has zero pressure where as the pressure of Dark Energy is negative. The status of Modified Chaplygin Gas is not something different but its legitimacy of being most favoured candidate has been being gradually strong in persuasion of the ongoing active research [7,- 10]. It should be marked that recently when it is attempted to understand the basic physics behind their peculiar indentities on the basis of Bose - Einstein Condensation of some boson fields without showing finger to particular candidate [11, -12 ], again a favourable result emerges in support of Chaplygin gas. The postulated condensate for understanding the mysterious physics, if endowed with negative pressure, is found to obey the exotic fluid equation folllowed by Chaplygin gas [13]. Thus, the research on Chaplygin gas, more correctly to say, if considered most favoured observational weightage, on Modified Chaplygin Gas deserves a special attention.\\

\noindent The pursuit of exploring the unknown character of Dark Energy becomes essentially entangled with other pertaining questions:\\

(i) The major problem to understand Einstein equation is the arbitrariness of energy - momentum tensor $T_ {\mu\nu}$. In the context of Dark Energy, as the energy source is in mysterious state, the problem is more severe. The general methodology adopted here is to impose some energy condition and to check its validity and violation. So, the work in this direction sheds light in exploration of new physics in different regimes defined on applied energy condition [14,- 40].\\

(ii) There are some parameters in Dark Energy model whose parameter spaces are still not well specified, and hence they are to be considered on phenomenological ground. Here, thermodynamical approach lends helping hand to impose some constraints on parameter space. Moreover, as the temperature generally incorporated in the thermodynamical analysis of Dark Energy is horizon's temperature the analysis offers new insights on spacetime, - and in long run on quantum mechanical interpretation of General Theory of Relativity [41].\\

\noindent In our previous publications [42, 43], we considered Modified Chaplygin Gas for the validity - check of Generalised Second Law of thermodynamics. There, we became able to derive a new expression for scale factor following full dynamics of the model. The present paper is greatly extended work of the previous ones. In addition to providing new results, the evolution of Modified Chaplygin Gas model is also analysed with interaction term which was previously lacking. This follow- up gives us the nature of variability of state parameter, $w$, - the most mysterious parameter of present - day cosmology.\\

\noindent The present work is organised as follows: Very briefly the main results of our previous works are presented in section II entitled as preliminaries. The section III contains our present works. The last section covers the summary and conclusion.\\

\section {Preliminaries}
 
\noindent Incorporating Modified Chaplygin Gas model given as

\begin{equation}
 p = B \rho - \frac{A}{\rho ^ {\alpha}},\; \;  A > 0, \; \; 0 < \alpha < 1
\end{equation}

\noindent in continuity equation

\begin{equation}
\dot{\rho} + 3 H \left ( \rho + p\right ) = 0
\end{equation}
\noindent we have:\\

\noindent (i) For specific stae parameter, $w =- 1.06 $ [44], that is, for the phantom case, the scale factor
\begin{equation}
a = \left[ \frac{3}{28} + \frac{25}{14}\left\lbrace (0.56a_*^{3(1+\alpha)/2} - 0.06)^{\frac{1}{\alpha + 1}} + C(t - t_*)\right\rbrace ^{(1+\alpha)}\right] ^{\frac{2}{3(1+\alpha)}}
\end{equation}

Here  $C^2=\frac{9}{4} \left( \frac{8\pi} {3 {M_p}^2} (2)^{\frac{1}{\alpha+1}} \rho_0 \right)$ \\

\noindent  And using Gibb's equation, it was obtained\\

\begin {equation}
T ds = - \frac{4 \pi R_H ^2}{4 \pi G} \dot{H} \left( \frac{3}{2 A}\right)  a ^ {\frac{1}{2}}
\end {equation}

\noindent If fluid temperature $T = T_H = \frac{1}{2\pi R_H}$, the horizon temperature,  we have

\begin {equation}
 ds = - \frac{2 \pi R_H ^3}{G} \dot{H} \left( \frac{3}{2 A}\right)  a ^ {\frac{1}{2}}
\end {equation}

\noindent (ii) For unspecific $w$, respective expressions appear as

\begin{eqnarray}
\lefteqn a \;\;  = {\left[ -\frac{(1 + w_0)}{n} + \frac{1}{n}\left\lbrace \left( n a_*^{3(1 + \alpha) (1+ N + w)} + (1 + w_0)\right) ^{\frac{1}{1 +\alpha}}\right . \right .} \nonumber \\
+ {\left. C(t - t_*)\right\rbrace ^{(1 + \alpha)}\left. \right]} ^{\frac{1}{3(1 +\alpha)(1 + N +w)}}
\end{eqnarray}

\noindent where  $ C^2 = 9 (1 + N + w)^2 \left\lbrace \frac{8\pi}{3M_p^2}(\frac{1}{1 + N +w})^{\frac{1}{1 + \alpha}} \rho_0\right\rbrace $\\

\begin{eqnarray}
Tds = - \frac{4 \pi R_H^2} {4 \pi G} \dot{H} \left[ dR_H - \frac{N + w +2}{(1 + N + w)} \left\lbrace a^{N + w + 2} - a_*^ {N + w +2}\right\rbrace \right] 
\end{eqnarray}

\section {Present Work :}

\subsection{Bound on $\alpha$}

\noindent It is observed in the equation (5) and equation (7) that the validity of Generalised Second Law rests on the signature of $\dot{H}$. The restriction imposed by $\dot{H}$ can be traced back to the constraint on Modified Chaplygin Gas model via the constraint on parameter space of $\alpha$. We have the expression

\begin {equation}
 \dot {H} = \frac{\ddot{a}}{a} - \left (\frac{\dot{a}}{a}\right ) ^2
\end {equation}

\noindent So the positivity or negativity of $\dot{H}$ rests on whether $a \ddot{a} > \left(\dot{a}\right) ^2 $ or  $a \ddot{a} < \left(\dot{a}\right) ^2 $  respectively. But unless we do some compromise in evaluation of $\dot{a}$ (of course without harming the basic physics involved), it is hard to be arrived at some illuminating expression. So, setting $a_* = 1$ and $ t = t_*$ (that is our compromise), we evaluate $a \ddot{a} < \left(\dot{a}\right) ^2 $. For the specific $w$, that is, $w = -1.06$, we obtain

\begin{equation}
 25 \alpha^2 + 354 \alpha + 123 < 0
\end{equation}

\noindent And for the unspecific $w$, following all the similar mathematical steps, we get

\begin{equation}
 \alpha > \frac{-2 - 3 K}{3 \left( K + 1\right)}
\end{equation}

\noindent $ K = \frac{C}{3 n \left(1 + N +w\right)} a^2 \left(1 + n + w\right)^\alpha$\\

\subsection{Interacting Modified Chaplygin Gas}

\noindent The mysterious state parameter $w \left(= \frac{p}{\rho}\right)$ is not a constant. Its different magnitude determines whether the concerned Dark Energy candidate is quintessence, phantom or cosmological constant. Obviously, the most pertinent questions are: Why is that variation? What is the nature of variation? One of the answers is the presence of some interaction between cold Dark Matter and Dark Energy. Interaction is incorporated into the continuity equation by some coupling term Q, and then detail of the investigation can be pursued.\\

\noindent In this respect, Chaplygin Gas model has an extra advantage, - as the model manifests both the identities, - the Dark Matter and Dark Energy provided with the specific conditions satisfied. So, within this model the proposed interaction is more conducive. Here, we consider the coupling term from Izquierdo and Pav$\acute{o}$n's work [45]. Only the coupling term proportional to Dark Energy density has been taken into account for analysis. In this case, coupling term Q is provided as $ Q = 3 \epsilon H \rho$, where $\epsilon$, the strength of interaction is non-negative and small. Now the continuity equation becomes

\begin{equation}
 \dot{\rho} + 3H \left(\rho + p\right) = -3 \epsilon H \rho
\end{equation}

\begin{equation}
 \Rightarrow \dot{\rho} + 3 H\left[\rho\left(1 + \epsilon\right) + p\right] = 0
\end{equation}

\noindent For $p$, if we consider the expression (1), and take $w = - 1.06$, finally it is obtained

\begin{equation}
 \rho^ {\alpha +1} = \frac{1}{\epsilon + 0.5 } \left[ 0.56 + \frac{\epsilon - 0.06}{a ^{3(\alpha +1) (0.5 + \epsilon)}}\right] {\rho_0}^{\alpha + 1}
\end{equation}

\noindent And if non-specific $w_0$ is retained, the expression reads as

\begin{equation}
 \rho^ {\alpha +1} = \frac{1}{1 + n + \epsilon + w_0} \left[ n  + \frac{1 + \epsilon + w_0}{a ^{3(\alpha +1) (1 + n + \epsilon + w_0)}}\right] {\rho_0}^{\alpha + 1}
\end{equation}

\noindent Further, we can assume 

$$ a = n {\rho_0}^{\alpha + 1} \equiv N {\rho}^ {\alpha +1}$$

\noindent which gives

$$\frac{n}{N} = \frac{{\rho}^{\alpha + 1}}{{\rho_0}^{\alpha +1}}$$

\noindent After some simple simplification, it is obtained

$$ a^{3(\alpha + 1) (1 +N + w)} = \frac{N (1 + w_0)}{n[n + (1 + w_0) - N]}$$

\noindent Taking logarithm on both sides, and performing the simple steps, we have

$$w = \frac{log [N(1 + w_0)] - log \left[n\lbrace n + (1 + w_0) - N\rbrace\right]}{3 (\alpha + 1) log a}$$

\begin{equation}
 w = \frac{K_1}{log a} - K_2
\end{equation}

\noindent where $K_1 = \frac{log [N(1 + w_0)]}{3 (\alpha + 1)}$  and  $ K_2 = \frac{log \left[n\lbrace n + (1 + w_0) - N\rbrace\right] }{3 (\alpha + 1)}$\\

\section{Summary and Conclusion}

\noindent Probably the compromise done at the level of $\dot{a}$, the expression (9) does not provide, at the first sight, any illuminating result regarding the parameter space of $\alpha$. Even it indicates contrary to our model concerned. But instead of inequality, if we consider equality sign, it means $\dot{H} = 0$. Physical interpretation of this is that there is no change of entropy. So in this situation, for these magnitudes of $\alpha$, either the two systems, - the fluid and the horizon are thermodynamically isolated or their mutual interaction is somehow shielded. This conclusion might have some linkage with the comment [15] made on the basis of apparent violation of Generalised Second Law that Chaplygin gas may have some upper bound in density at the initial state. Not only the specific magnitude of $w$,  for the unspecific $w$ also the same interpretation prevails with the expression (10). In this case, another subtle point comes into the picture. The pressence of the constant C in (10) where 

$$ C^2= 9 (1 + N + w)^2 \frac{8\pi} {3 {M_p}^2} \lbrace\frac{1}{1 + n +w_0}\rbrace^{\frac{1}{\alpha + 1}} \rho_0 $$

\noindent for unspecific $w$, shows some impact of energy density on $\alpha$. For various findings, the question of energy density should be handled very cautiously [46, - 50].\\

\noindent In interacting Modified Chaplygin Gas scenario, the expression (13) shows that there is no variation of density at $\epsilon = 0.06$ for $w_0 = -1.06 $, and for unspecific $w_0$ also there is similar situation at $ w_0 = - (1 + \epsilon)$ as shown by the expression (14). This result is an indication of crucial impact of interaction strength on Dark Energy evolution. From the expression (15) we get the clear-cut dependency of $w$ on scale factor $a(t)$.\\

\noindent {\bf{Acknowledgement}}\\

\noindent This work has been finalised by one (*) of the authors staying at SINP, KOLKATA as an IASc-INSA-NASI (teacher) summer research fellow against PHYT80, 2010.\\

\noindent {\bf{References}}\\

\noindent [1] Perlmutter, S. J. et al. (1999), Astrophys. J. {\bf{517}}, 565; (1998) astro-ph/9812133\\

\noindent [2] Riess, A.G. et al. (1998), Astrophys. J. {\bf{116}}, 1009, 
\noindent Riess, A.G. et al. (2004), Astrophys. J. {\bf{607}}, 665; astro-ph/0402512\\

\noindent [3] Garnavich,P.M. et al. (1998), Astrophys. J. {\bf{509}}, 74 \\

\noindent [4] Spergel, D.N.  et al. (2003), Astrophys. J. Suppl. {\bf{148}}, 175; astro-ph/9812133 and references therein\\

\noindent [5] Tonry, J.L. et al. (2003), Astrophys. J. {\bf{594}}, 1 \\

\noindent [6] Wang, Y. and Tegmark, M. (2004) Phys. Rev. Lett. {\bf{92}}, 241302 \\

\noindent [7] Debnath, U., and  Chakrabarty, S., (2006), gr-qc/0601049\\

\noindent [8] Kamenchichik, A., Moschella, U., and Pasquier (2001), Phys.Lett. B {\bf{511}}, 265\\

\noindent [9] Benaoum, H.B., (2002), hep-th/0205140 ;\\ Bento, M.C., Bertolami, O., and Sen, A.A. (2002), gr-qc/0202064\\

\noindent [10] Jackiw, R.``(A Particle Field Theorist's ) Lecture on (Supersymmetric, Non-Abelian), Fluid Mechanics (and d- Branes)", physics/0010042\\

\noindent [11] Fukuyama, T., Morikawa, M., (2009), astro-ph/0905.0173\\

\noindent [12] Fukuyama, T., Morikawa, M., (2006), Prog. Theor. Phys. {\bf{115}} 1047; astro-ph/0509789\\

\noindent [13] Popov, V.A., (2009),  gr-qc/0912.1609\\

\noindent [14] Caldwell, R. R., (2002), Phys. Lett. B, {\bf{545}}, 23,
\noindent Caldwell, R.R.,  Kamionkowski, M. and Weinberg, N.N., (2003) Phys. Rev. Lett., 
{\bf{91}}, 071301,\\

\noindent [15]Izquierdo Germ$\acute{a}$n, Pav$\acute{a}$n Diego, (2006) Phy. Lett. B, {\bf{633}}, 420\\

\noindent [16] Overduin, J.M., Cooperstock, F.I., (1998), Phys. Rev.D {\bf{58}}, 043506\\

\noindent [17] Aremendariz-Picon, C., Damour, T. and Mukhanov, V., (1999), Phys. Lett. B {\bf{458}}, 209\\

\noindent [18] Chiba, T., Okabe, T. and Yamaguchi, M, (2000), Phys. Rev.D {\bf{62}}, 023511\\

\noindent [19] Sen, A., (2002), J. High Energy Phys. {\bf{04}}, 048;   (2002), {\bf{07}}, 065; (2002), Mod. Phys. Lett. A {\bf{17}}, 1799 \\

\noindent [20] Garousi, M.R.,  (2000), Nucl. Phys. B {\bf{584}}, 284;  (2003), J. High Energy Phys. {\bf{04}}, 027\\

\noindent [21] Singh, Parampreet, Sami, M.,Dadhich, N. (2003), Phys. Rev.D {\bf{68}}, 023522 and references therein\\

\noindent [22] Panda, S., Perez-Lorenzana, A. (2001), Nucl. Phys. B {\bf{584}}, 284\\

\noindent [23] Srivastava, S.K., (2004), gr-qc/040974\\

\noindent [24] Sami, M., Toporensky, A., Tretjakov, P.V., Tsujikawa, S (2005), Phys. Lett. B {\bf{619}}, 193 [hep-th/0504155]\\

\noindent [25] Calcagni, G., Tsujikawa, S., Sami, M.,  (2005), Class. Quan. grav. {\bf{22}}, 3977 [hep-th/0505193\\

\noindent [26] srivastava, S. K.;(2005), Phys. Lett. B {\bf{619}}, [astro-ph/ 0407048]\\

\noindent [27] Jackiw, R, (2000), Physics/0010042\\

\noindent [28] Bartolami, O. et al, (2004), Mon. Not. R. Astron. Soc.{\bf{353}}, 329 [astra.phy/0402387].\\

\noindent [29] Bento, m. C., Bertolami, O., Sen, A. A., (2002), Phys. Rev.D {\bf{66}} 043507 [gr-qc/0202064]\\

\noindent [30] Capozziello, S., (2002), Int. J. Mod. Phy D {\bf{11}}483\\

\noindent [31] Capozziello, S., Carloni, S., Troisi, A.,(2003), astro-ph/0303041\\

\noindent [32] Caroll, S. M., Duvvuri, V., Trodden, M., Turner, M. S., (2004), Phys. Rev.D {\bf{70}}, 043528, astro-ph/0306438\\

\noindent [33] Dolgov, A. D., Kawasaki, M., (2003), Phys. lett. B {\bf{573}} 1; astro-ph/0307285.\\

\noindent [34] Soussa, m. E., Woodard, R. P., (2004),Gen. Relativ. Gravit.  {\bf{36}} 855, astro-ph/0308114.\\

\noindent [35] Nojiri, S, Odintsov, S. D., (2004), Phys. lett. A {\bf{19}} 627; hep-th/0310045.\\

\noindent [36] Nojiri, S, Odintsov, S. D., (2003), Phys. Rev. D {\bf{68}} 123512, hep-th/0307228\\

\noindent [37] Abdalaa, M. C. B., Nojiri, S., Odintsov, s. D., (2005), Class. Quan. Grav. {\bf{22}}, 35, [hep-th/0409117]\\

\noindent [38] Mena , O., Santiago, J., Weller, J., (2006), Phys. Rev.Lett. {\bf{96}}, 041103\\

\noindent [39] Atazadeh, K., Sepangi, H.R. (2006), gr-qc/0612135\\

\noindent [40] Bouhmadi-Lopez, M. et al, (2005),   astro-ph/0512124; (2006), gr-qc/0612135; (2007), arXiv:0707.2390; (2007) arXiv: 0706.3896 [astro-ph]\\

\noindent [35] Wang, B., Gong, Y., Abdalla, E. (2006), Phys. Rev.D. {\bf{74}}, 0083520\\

\noindent [36] Padmanabhan, T.,(2010), Rep. Prog. Phys. {\bf{73}}, 046901; gr-qc/0911.5004 and references therein\\

\noindent [37] Hawking, S. W. and   Ellis, G. F. R., (1980), The large scale structure of space-time (Cambridge,
Cambridge University Press, 1980).\\

\noindent [38] Bekenstein, J., (1973), Phys. Rev.D. {\bf{7}}, 2333\\

\noindent [39] Hawking, S. W. (1975), Commun. Math. Phys. {\bf{43}}, 199\\

\noindent [40] Gibbons, G.W., Hawking, S. W., (1977), Phys. Rev.D. {\bf{15}}, 2738\\

\noindent [41] Padmanabhan, T.,(2010), Rep. Prog. Phys. {\bf{73}}, 046901; gr-qc/0911.5004 and references therein\\

\noindent [42] Dev choudhury, B. K., Saikia, J., (2009), Proc. 31st ICRC, Lodz, Poland {\bf{H.E. 2.3}}, astr-ph/0906.0644\\

\noindent [43] Saikia, J., Dev choudhury, B. K., (2009), Proc. 31st ICRC, Lodz, Poland {\bf{H.E. 2.3}}, astr-ph/0906.0646\\

\noindent [44] Knop, R.A., et al, (2003), Astrophys. J. {\bf{598}} 102;\\
\noindent Melchiorri, A., (2004), ``Plenary Talk given at Exploring the Universe, Moriond, astr-ph/0406652\\

\noindent [45] Izquierdo, G., and Pav$\acute{o}$n, D., (2010), astro-ph/1004.2360\\

\noindent [46] Ford, L.H., and Roman, T.A., (2001), gr-qc/0009076\\

\noindent [47] Babichev, E., DokuChaev, V., Eroshenko, Yu., (2004), Phys. Rev. Lett. {\bf{93}} 021102\\

\noindent [48] Nojiri, S., Odintsov, S.D., (2004), Phys. Rev. D {\bf{70}} 103522\\

\noindent [49] Ford, L.H., (1978), Proc. Roy. Soc. Lond. A {\bf{364}} 227\\

\noindent [50] Ford, L.H., (1991), Phys. Rev. D {\bf{43}} 3972\\

\end{document}